\documentclass[10pt,oneside,a4paper]{article}

\usepackage{physics}
\usepackage{graphicx}
\usepackage{amssymb}
\usepackage{xcolor}
\usepackage{authblk}

\date{}

\begin{document}

\title{Impact of Parallel Gating on Gate Fidelities in Linear, Square, and Star Arrays of Noisy Flip-Flop Qubits}

\author{Marco De Michielis\footnote{marco.demichielis@cnr.it} , Elena Ferraro}
\affil{CNR-IMM, Unit of Agrate Brianza, Via C. Olivetti 2,\\ 20864 Agrate Brianza (MB), Italy}

\maketitle

\begin{abstract}
Successfully implementing a quantum algorithm involves maintaining a low logical error rate by ensuring the validity of the quantum fault-tolerance theorem. The required number of physical qubits arranged in an array depends on the chosen Quantum Error Correction code and the achievable physical qubit error rate. As the qubit count in the array increases, parallel gating —simultaneously manipulating many qubits— becomes a crucial ingredient for successful computation.

In this study, small arrays of a type of donor- and quantum dot-based qubits, known as flip-flop qubits, are investigated. 
Simulation results of gate fidelities in linear, square and star arrays of four flip-flop qubits affected by realistic 1/f noise are presented to study the effect of parallel gating. The impact of two, three and four parallel one-qubit gates, as well as two parallel two-qubit gates, on fidelity is calculated by comparing different array geometries.
Our findings can contribute to the optimized manipulation of small flip-flop qubit arrays and the design of larger ones.
\end{abstract}

\section{Introduction}

In the realm of semiconductor qubit holders based on donor atoms and quantum dots in silicon \cite{Vandersypen-2017,McCallum-2021,Burkard-RevModPhys2023,DeMichielis-JPhysD-2023}, the flip-flop (FF) qubit consists of a $^{31}$P donor atom situated within a $^{28}$Si bulk, positioned at a given distance from the Si/SiO$_2$ interface. An electric field, generated by a metal gate atop the SiO$_2$ layer, controls the movement of the donor-bound electron between the donor site and the Si/SiO$_2$ interface region. 
By changing the applied electric field the antiparallel electron-nuclear spin states, which defined the FF qubit, can be manipulated.
The FF qubit was presented by Tosi \textit{et al.} in 2017 \cite{Tosi-2017}, gained increasing attention in the following years \cite{Tosi-2018,Morello-2020,Ferraro-2022,Rei-2022,Calderon-2022} and was experimentally demonstrated by Savytskyy \textit{et al.} in 2023 \cite{Savytskyy2023Science}.

FF qubits are of interest due to their exploitation of long-range dipole-dipole interactions \cite{Tosi-2017}, which can mitigate the stringent requirements for precise qubit placement and the resulting inter-qubit spacing. 

A reliable assessment of the increase in gate infidelity induced by parallel gating needs to be considered for an accurate estimation of errors in Quantum Error Correction codes, enabling fault-tolerant quantum computation.

When considering array geometry, in addition to the one-dimensional linear array (LA) and the two-dimensional square array (SA), a second two-dimensional array has to be considered: the hexagonal array. 
At a given inter-qubit distance, the LA is easy to fabricate due to its low qubit density, providing free area for initialization/readout devices, but it is not suitable for very large arrays, unless segmented. The SA is more complex to fabricate due to its higher qubit density and provides less space for initialization/readout devices, but it should be preferred for large arrays due to its 2D geometry. The hexagonal array offers an interesting trade-off, with a lower density than the square one, making it still suitable for large-scale arrays.
Hexagonal lattices can be generated by composing replicas of a small star array (STA), where a central qubit is directly coupled with three equidistant qubits placed at the vertices of an equilateral triangle. 

The effects of two parallel one- and two-qubit gates applied to four noisy FF qubits arranged in a LA and a SA have been previously studied \cite{DeMichielis-AQT-2024}. Here, we extend this study to include results on a STA and to investigate the effect on fidelity of more than two parallel operations within each array.

\section{Model}

The Hamiltonian $\hat{H}_{A}$ describing an array of $N$ FF qubits is:
\begin{equation} \hat{H}_{A}=\sum_{i=1}^N\hat{H}^i(\Delta E_{z}^i,E_{ac}^i)+\sum_{i=1}^{N-1}\sum_{j=i+1}^N\hat{H}_{int}^{i,j}(\Delta E_{z}^i,\Delta E_{z}^j,r_{ij}),
\label{eq:H_A}
\end{equation}
where $\hat{H}^i$ is the Hamiltonian of the single FF qubit and $\hat{H}_{int}$ is the dipole-dipole interaction term between qubits $i$ and $j$ \cite{DeMichielis-AQT-2024}. The Hamiltonian $\hat{H}^i$ is a function of a DC electrical field difference $\Delta E_z^i=E_z^i-E_z^0$, where $E_z^i$ is the imposed vertical electric field and $E_z^0$ represents the vertical electric field at the ionization point, i.e. the point at which the electron is shared halfway between the P donor and the interface, and of an AC electrical field $E_{ac}^i$. Both are used to manipulate the qubit. The interaction Hamiltonian $\hat{H}_{int}^{i,j}$ depends on the DC electrical field difference in qubit $i$ and in qubit $j$, and it is proportional to $r_{ij}^{-3}$, where $r_{ij}$ is the distance between the two qubits \cite{Tosi-2017,DeMichielis-AQT-2024}.

In the one-dimensional array model only the first neighbour inter-qubit interactions are included thus $\hat{H}_{int}^{i,j}$=0 for $j$$>$$i+1$, whereas in the other two two-dimensional arrays all inter-qubit interactions are considered. 

\begin{figure}[htbp!]
\centering
    a) \includegraphics[width=0.24\linewidth]{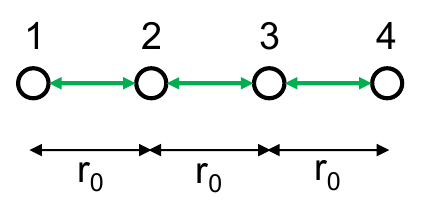}       
    b) \includegraphics[width=0.162\linewidth]{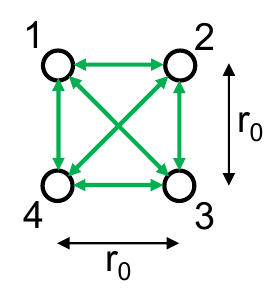}       
    c) \includegraphics[width=0.276\linewidth]{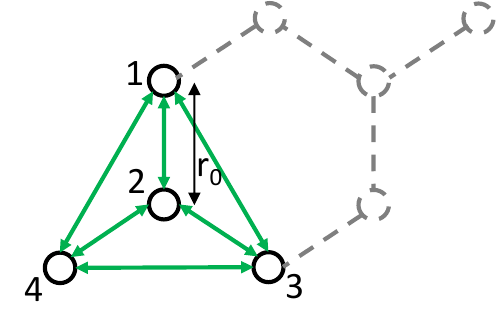}       
    \caption{a) LA scheme composed by four FF qubits equally displaced, where $r_0$ is the inter-qubit distance. Only the first neighbour inter-qubit interactions, highlighted in green, are included in the model. b) SA scheme composed by four FF qubits where all inter-qubit interactions are included. c) STA scheme composed by four FF qubits where all inter-qubit interactions are included. The addition of a rotated replica of the STA leads to an hexagonal lattice as highlighted by the dashed gray lines.}
    \label{fig:schemesAndInterac}
\end{figure}

We simulate the case corresponding to $N$=4, with the qubits arranged in a LA, in a SA and in a STA as depicted in Figure \ref{fig:schemesAndInterac}. 

The STA can be described as a central qubit directly coupled with three equidistant qubits placed at the vertices of an equilateral triangle. The shortest inter-qubit distance $r_0$ is the same in all three arrays and is set to 360 nm. 

The quantum gates under investigation here, namely the one-qubit gates $R_z(-\frac{\pi}{2})$ and $R_x(-\frac{\pi}{2})$ and the two-qubit gate $\sqrt{iSWAP}$, constitute a universal set of quantum gates achievable through total electrical manipulation \cite{Tosi-2017}, applying the control sequences given by the $\Delta E_{z}(t)$ and $E_{ac}(t)$ signals reported in \cite{DeMichielis-AQT-2024}.

In order to study the effect of multiple parallel operations, we apply simultaneously the one-qubit operations to two, three or four qubits in the array, and the two-qubit operations on four qubits. Each one of these cases, referred to as configurations, is labeled as reported in Table \ref{tab:configurationsLA}.

\begin{table}[htbp!]
\centering
\caption{Correspondence between configurations and types of operation in the arrays.}
\resizebox{\textwidth}{!}{
    \begin{tabular}{|l|l|}
      \hline
      Configuration  & Type of Operation  \\
      \hline 
      ci ($i=1, \dots, N$)   & One-qubit operation on qubit $i$ while the others are idle \\
      \hline
      cij ($j\neq i=1, \dots, N$)   & Two parallel one-qubit operations or single two-qubit operation on qubits $i$ and $j$\\
      & while the others are idle\\
      \hline
      cijk ($k\neq j\neq i=1, \dots, N$)   & Three parallel one-qubit operations on qubits $i$,$j$ and $k$ while the other is idle \\
      \hline
      cijkl ($l\neq k\neq j\neq i=1, \dots, N$)   & Four parallel one-qubit operations applied on all the qubits $i$,$j$,$k$ and $l$\\
      \hline
      cij-kl ($l\neq k\neq j \neq i=1, \dots, N$) & Two parallel two-qubit operations on the qubit couples $ij$ and $kl$ \\
      \hline
    \end{tabular}}
    \label{tab:configurationsLA}   
\end{table}

The figure of merit chosen to compare the different arrays is the entanglement fidelity $F$ \cite{DeMichielis-AQT-2024, nielsen1996entanglement}. 
Following  \cite{DeMichielis-AQT-2024} for each gate under study, one hundred instances of the charge noise with 1/f spectrum, ranging from $f_{min}$=50 kHz to $f_{max}$=22 GHz, are generated in the time domain with an amplitude $\alpha_{\Delta E_{z}}$  and added to the ideal sequence signals performing the operation for each qubit. Each qubit in the array is affected by a different noise instance because noise correlations between qubits are not considered. Finally, we take the average over the resulting entanglement infidelities. 

\section{Results}
In this Section, the results of simulated infidelities for parallel one-qubit and two-qubit operations are presented when the noise amplitude $\alpha_{\Delta E_z}$ spans a range from 0 to 100 V/m. For the one-qubit case, after presenting the outcome of a single operation, the results on infidelities for two, three, and four parallel gates are reported. Then, simulation results for one and two parallel two-qubit operations are shown.

\subsection{One-qubit Operations}
\begin{figure}[htbp!]
\centering
    a)\includegraphics[width=0.3\linewidth]{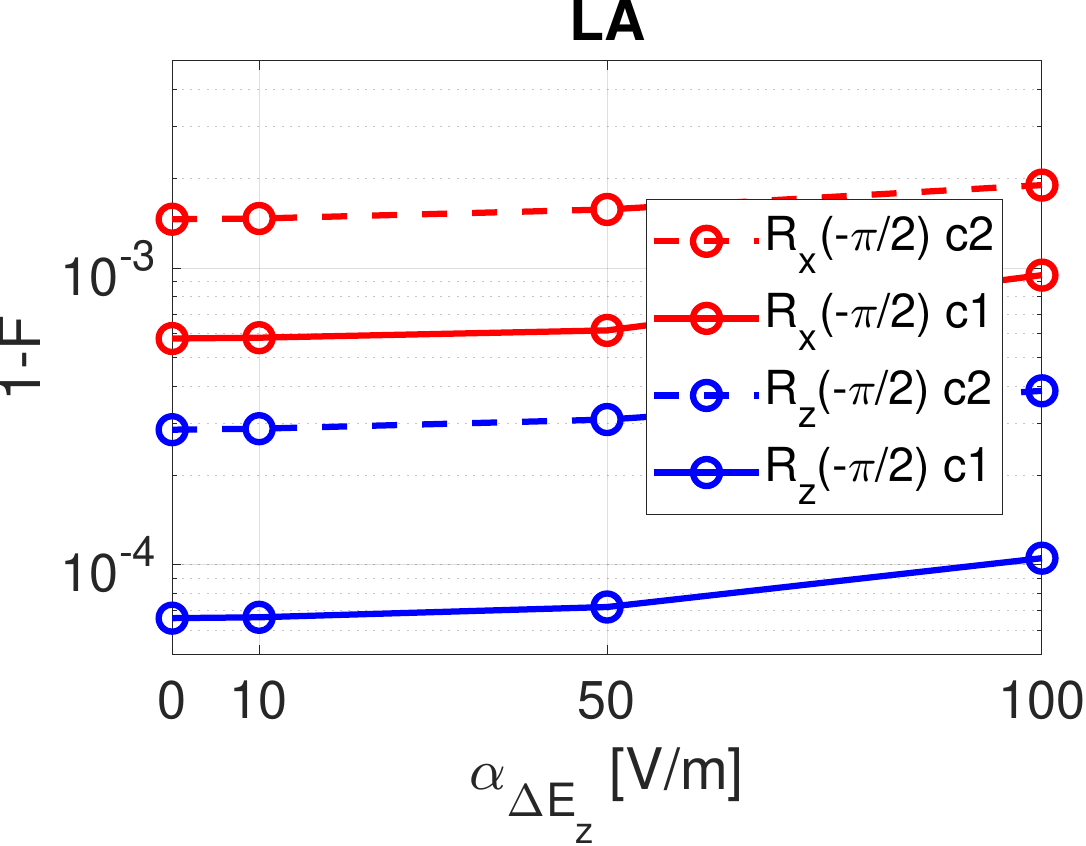}
    b)\includegraphics[width=0.3\linewidth]{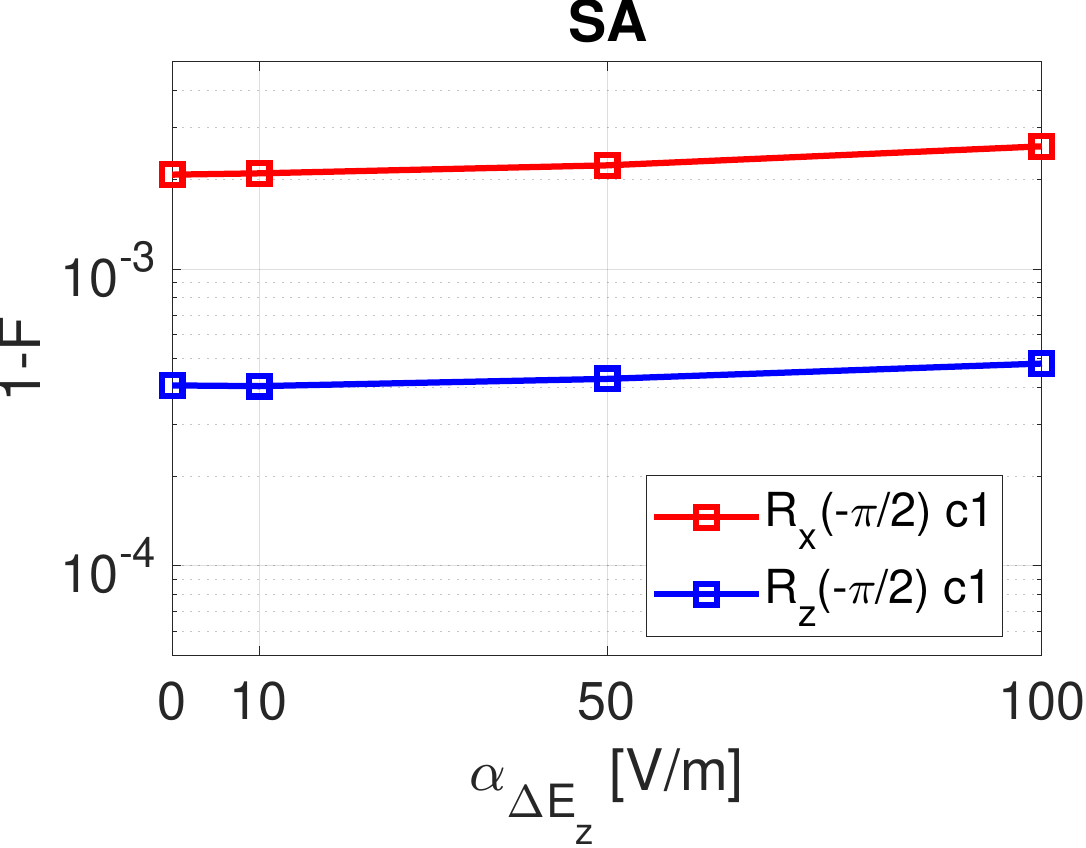} 
    c)\includegraphics[width=0.3\linewidth]{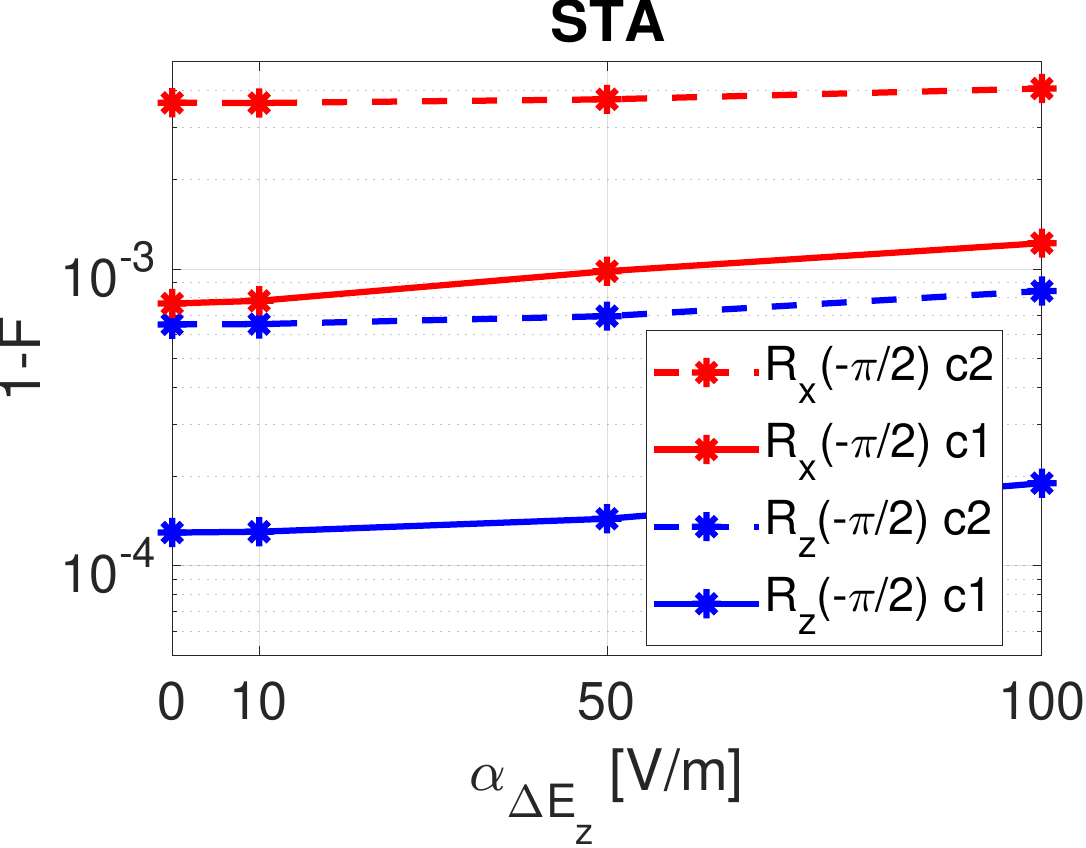}   
    \caption{$R_{z}(-\frac{\pi}{2})$ (blue curves) and $R_{x}(-\frac{\pi}{2})$ (red curves) infidelities versus the noise amplitude $\alpha_{\Delta E_{z}}$ for a) the c1 and c2 configurations in the LA, b) the c1 configuration in the SA and c) the c1 and c2 configurations in the STA.}
    \label{fig:comparison1qOp}
\end{figure}

Figure \ref{fig:comparison1qOp} shows the infidelity $1-F$ as a function of the noise amplitude $\alpha_{\Delta E_z}$ when a one-qubit gate, $R_z(-\frac{\pi}{2})$ and $R_x(-\frac{\pi}{2})$, is performed on one qubit while the others are in an idle state in a LA (a), in a SA (b) and in a STA (c). The configurations of interest are the c1 and c2 configurations for the LA, the c1 for the SA and the c1 and c2 for the STA. All other possible configurations are geometrically equivalent to the ones presented and thus yield the same results for infidelities. This consideration holds from this point onward, regardless of the type of operations performed. 
All the gate infidelities increase as the noise amplitude is raised, with the infidelity of $R_x(-\frac{\pi}{2})$ being higher than that of $R_z(-\frac{\pi}{2})$.

In correspondence to a realistic noise amplitude, i.e. $\alpha_{\Delta E_z}=$ 50 V/m, we report in Table \ref{tab:one} the best and the worst infidelity values for each operation and for each array type, specifying in parenthesis the corresponding configuration.

\begin{table}[htbp!]
\centering
    \caption{One-qubit operations best an worst infidelity values when $\alpha_{\Delta E_z}$= 50 V/m, in parenthesis the corresponding configuration is specified.}
    \begin{tabular}{|c|c|c|c|}
      \hline
       One-qubit &    \multicolumn{3}{c|}{1-F} \\
       \cline{2-4}
            operation      &  LA &  SA &  STA  \\       
      \hline 
       $R_z(-\frac{\pi}{2})$ &  $7.2\cdot 10^{-5}$ (c1)&  $4.3\cdot 10^{-4}$ (c1)&$1.4\cdot 10^{-4}$ (c1)\\ 
       &   $3.1\cdot 10^{-4}$ (c2)& - &  $6.9\cdot 10^{-4}$ (c2)\\
      \hline 
      $R_x(-\frac{\pi}{2})$   &  $6.2\cdot 10^{-4}$ (c1)&  $2.2\cdot 10^{-3}$ (c1)&  $9.8\cdot 10^{-4}$ (c1)\\ 
       &  $1.6\cdot 10^{-3}$ (c2)& - &  $3.7\cdot 10^{-3}$ (c2)\\
      \hline
    \end{tabular}
    \label{tab:one}
\end{table}

The c1 configuration always shows the lower infidelity value in both $R_z(-\frac{\pi}{2})$ and $R_x(-\frac{\pi}{2})$ operations because it corresponds to a qubit that is less disturbed by the idle surrounding qubits. The LA is the most favorable geometry for performing single-qubit operations.

\subsubsection{Two Parallel One-qubit Operations} 
\begin{figure}[htbp!]
\centering
    a)\includegraphics[width=0.3\linewidth]{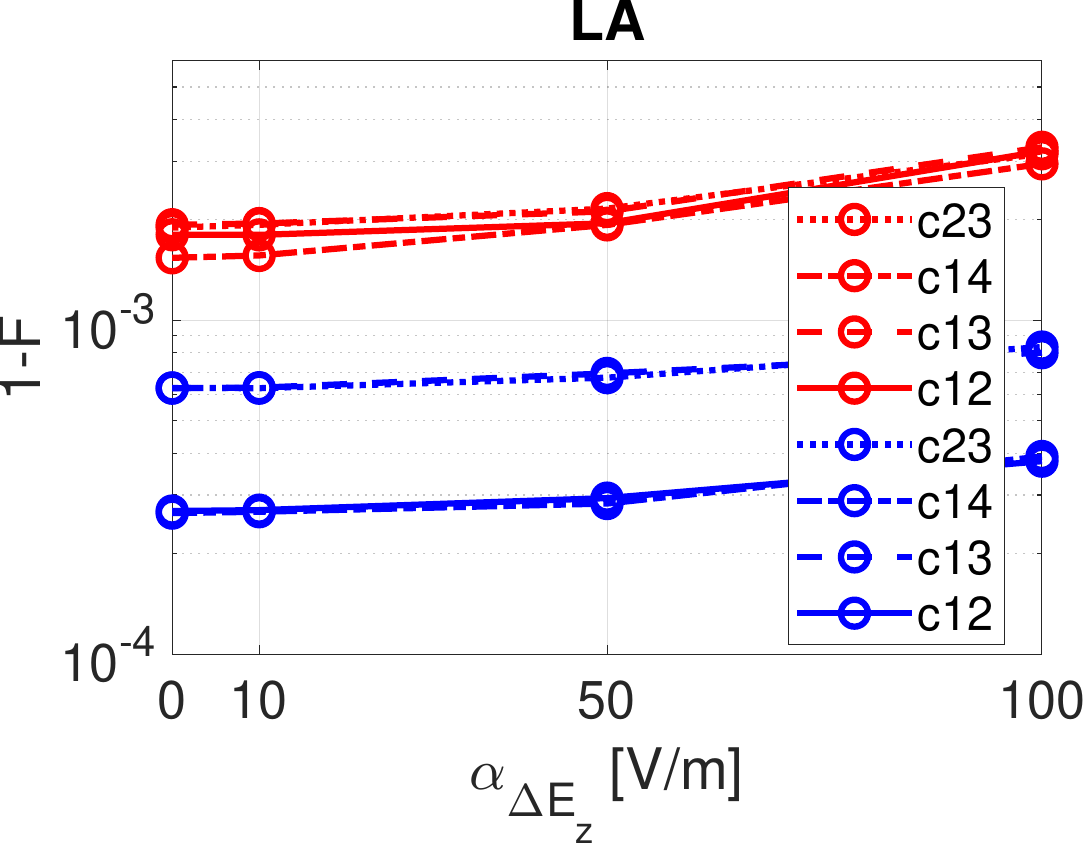}
    b)\includegraphics[width=0.3\linewidth]{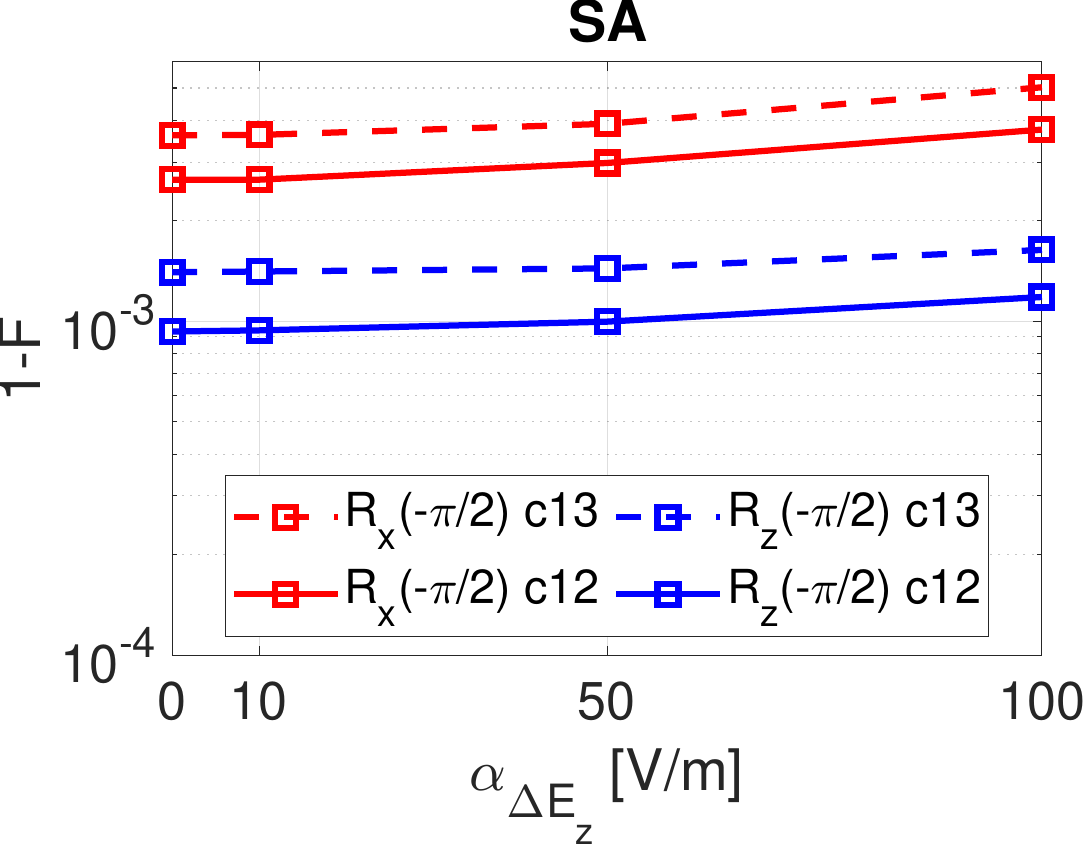} 
    c)\includegraphics[width=0.3\linewidth]{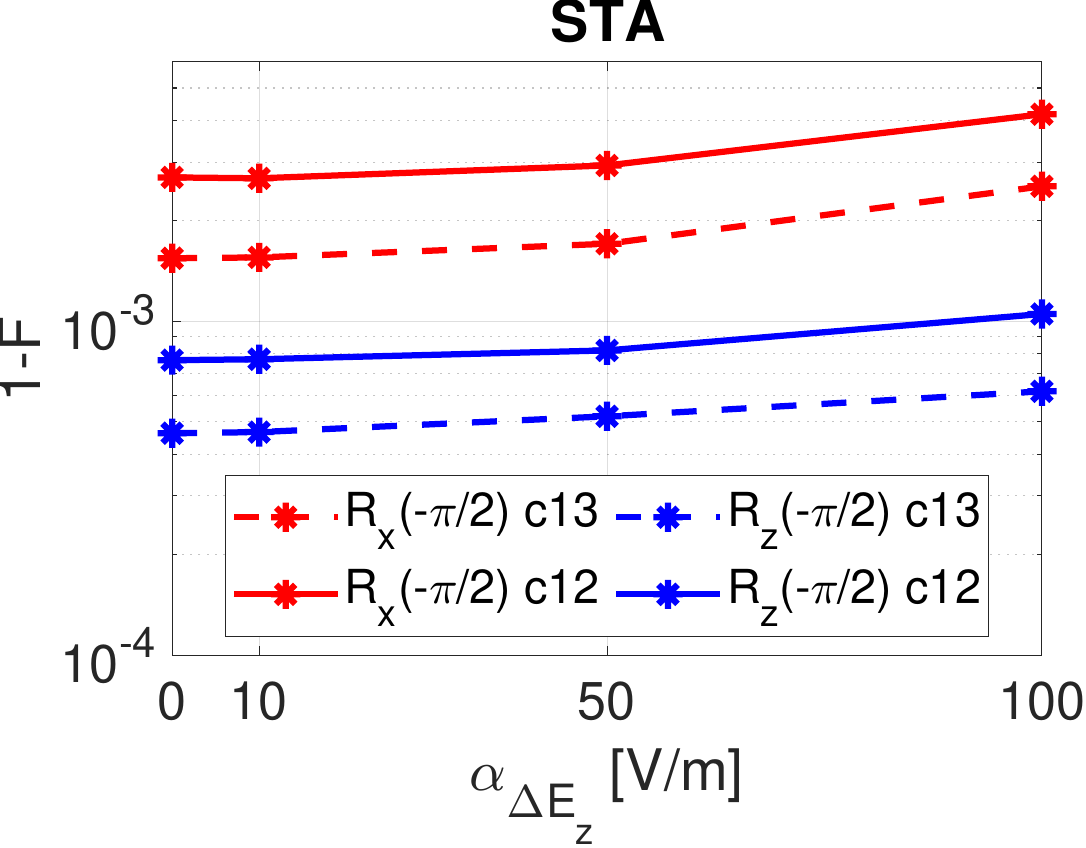}   
    \caption{Two parallel $R_{z}(-\frac{\pi}{2})$ (blue curves) and $R_{x }(-\frac{\pi}{2})$ (red curves) infidelities versus the noise amplitude $\alpha_{\Delta E_{z}}$ for a) the c12, c13, c14 and c23 configurations in the LA, b) the c12 and c13 configurations in the SA and c) the c12 and c13 configurations in the STA.}
    \label{fig:comparisonTwo1qOp}
\end{figure}

Figure \ref{fig:comparisonTwo1qOp} illustrates the infidelity $1-F$ as a function of the noise amplitude $\alpha_{\Delta E_z}$ when two parallel one-qubit gate, $R_z(-\frac{\pi}{2})$ and $R_x(-\frac{\pi}{2})$, are applied to a pair of qubits while the others are in an idle state. The analyzed configurations depend on the array geometry. Here, we study all pair of qubits at distances $r_0$ (configurations c12 and c23), $2r_0$ (configuration c13) and $3r_0$ (configuration c14) for the LA (a). For the SA, we study pairs at distance $r_0$ and $\sqrt{2}r_0$ corresponding respectively to c12 and c13 (b). In the STA, we consider pairs at distance $r_0$ and $\sqrt{3}r_0$, corresponding respectively to c12 and c13 (c).  

We report in Table \ref{tab:two} the best and the worst infidelity values for each operation and for each array type when $\alpha_{\Delta E_z}=$ 50 V/m.

\begin{table}[htbp!]
\centering
    \caption{Two parallel one-qubit operations best an worst infidelity values when $\alpha_{\Delta E_z}$= 50 V/m, in parenthesis the corresponding configuration is specified.}
    \begin{tabular}{|c|c|c|c|}
      \hline
       Two parallel  &   \multicolumn{3}{c|}{1-F} \\
       \cline{2-4}
     one-qubit operations &   LA  & SA & STA  \\       
      \hline 
       $R_z(-\frac{\pi}{2})$ &  $2.8\cdot 10^{-4}$ (c14)& $1.0\cdot 10^{-3}$ (c12)& $5.2\cdot 10^{-4}$ (c13)\\ 
       &   $6.9\cdot 10^{-4}$ (c13)&$1.4\cdot 10^{-3}$ (c13) & $8.2\cdot 10^{-4}$ (c12)\\
      \hline 
      $R_x(-\frac{\pi}{2})$   &  $1.9\cdot 10^{-3}$ (c14) & $3.0\cdot 10^{-3}$ (c12)& $1.7\cdot 10^{-3}$ (c13)\\ 
       &  $2.2\cdot 10^{-3}$ (c23)& $3.9\cdot 10^{-3}$ (c13)& $2.9\cdot 10^{-3}$ (c12)\\
      \hline
    \end{tabular}
    \label{tab:two}
\end{table}

When two parallel one-qubit operations are performed, the number of possible configurations in the arrays studied increases, and we observed more variability in the infidelity values. The LA allows for a larger number of configurations to be considered, and the lowest infidelity values are obtained in the c14 configuration, where the operated qubits are at the extremes of the array. Conversely, higher infidelity values are found in the c13 and c23 configurations, although these configurations produce very similar results.

In the SA, the best infidelity values are obtained when the operated qubits are closer together (on the edge of the square) compared to configurations where they are farther apart (on the diagonal of the square). In the STA, it is preferable for the operated qubits to be on the edge of the triangle rather than having one qubit in the center, as the central qubit is much more affected by the disturbance from the idle qubits.

In general, we may conclude that the LA also provides better results for both one-qubit operations, even if the STA is preferred for the $R_x(-\frac{\pi}{2})$ operation.

\subsubsection{Three Parallel One-qubit Operations}
\begin{figure}[htbp!]
\centering
   a)\includegraphics[width=0.3\linewidth]{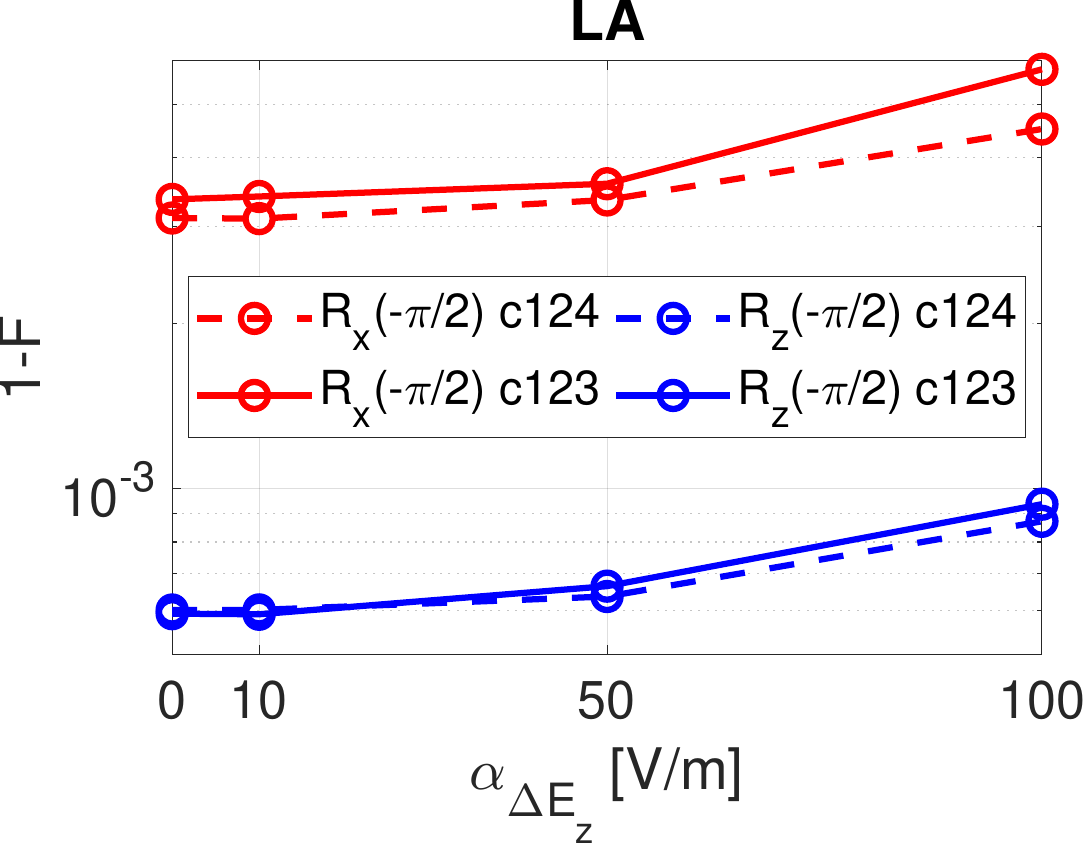}
   b)\includegraphics[width=0.3\linewidth]{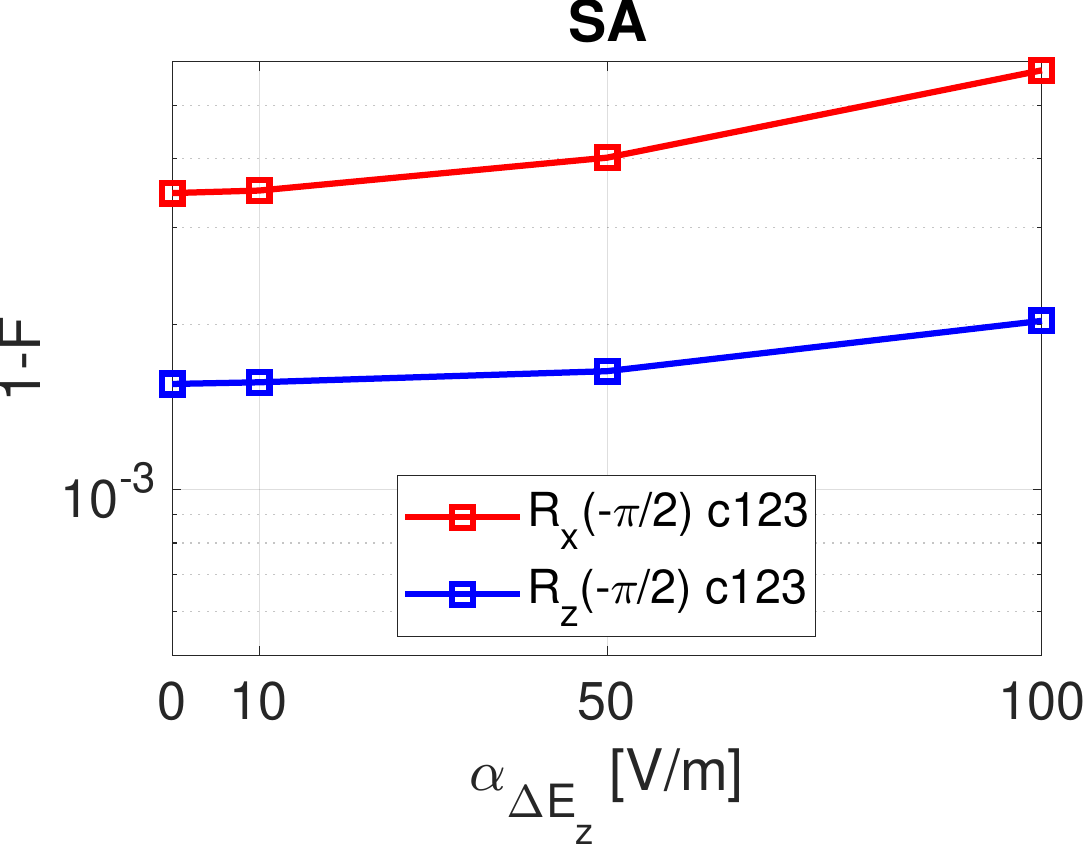} 
   c)\includegraphics[width=0.3\linewidth]{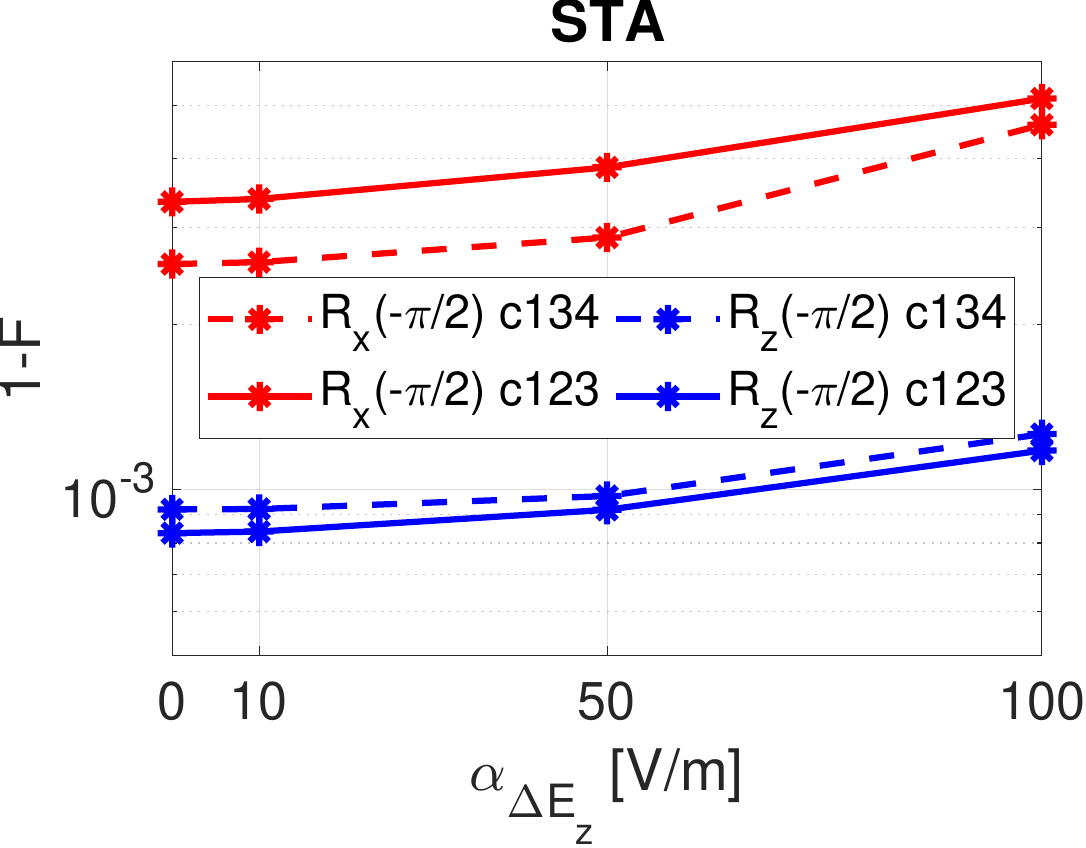}   
    \caption{Three parallel $R_{z}(-\frac{\pi}{2})$ (blue curves) and $R_{x }(-\frac{\pi}{2})$ (red curves) infidelities versus the noise amplitude $\alpha_{\Delta E_{z}}$ for a) the c123 and c124 configurations in the LA, b) the c123 in the SA and c) the c123 and c134 configurations in the STA.}
    \label{fig:comparisonThree1qOp}
\end{figure}

Figure \ref{fig:comparisonThree1qOp} illustrates the infidelity $1-F$ as a function of the noise amplitude $\alpha_{\Delta E_z}$ when three parallel one-qubit gates, $R_z(-\frac{\pi}{2})$ and $R_x(-\frac{\pi}{2})$, are applied to a triplet of qubits while the fourth qubit remains idle. The analyzed configurations are c123 and c124 for LA (a), c123 for SA (b) and c123, c134 for STA (c). 

Table \ref{tab:three} collects the best and worst infidelity results for all the operations and arrays studied when $\alpha_{\Delta E_z}=$ 50 V/m.

\begin{table}[htbp!]
\centering
    \caption{Three parallel one-qubit operations best an worst infidelity values when $\alpha_{\Delta E_z}$= 50 V/m, in parenthesis the corresponding configuration is specified.}
    \begin{tabular}{|c|c|c|c|}
      \hline
       Three parallel &   \multicolumn{3}{c|}{1-F} \\
       \cline{2-4}
       one-qubit operations &   LA  & SA & STA  \\       
      \hline 
       $R_z(-\frac{\pi}{2})$ &  $6.4\cdot 10^{-4}$ (c124)& $1.6\cdot 10^{-3}$ (c123)& $9.2\cdot 10^{-4}$ (c123)\\ 
       &   $6.6\cdot 10^{-4}$ (c123)& - & $9.7\cdot 10^{-4}$ (c134)\\
      \hline 
      $R_x(-\frac{\pi}{2})$   &  $3.3\cdot 10^{-3}$ (c124) & $4.0\cdot 10^{-3}$ (c123)& $2.9\cdot 10^{-3}$ (c134)\\ 
       &  $3.6\cdot 10^{-3}$ (c123)& - & $3.9\cdot 10^{-3}$ (c123)\\
      \hline
    \end{tabular}
    \label{tab:three}
\end{table}

The infidelity results for the different configurations considered yield similar outcomes in both the LA and the STA, while only one configuration for the SA is considered due to the symmetry of the array. For both operations, the SA shows larger infidelities compared to the STA and the LA. The LA configuration yields lower infidelity values for $R_z(-\frac{\pi}{2})$, while the STA configuration yields lower infidelity values for $R_x(-\frac{\pi}{2})$.

\subsubsection{Four Parallel One-qubit Operations}
\begin{figure}[htbp!]
\centering
   a)\includegraphics[width=0.3\linewidth]{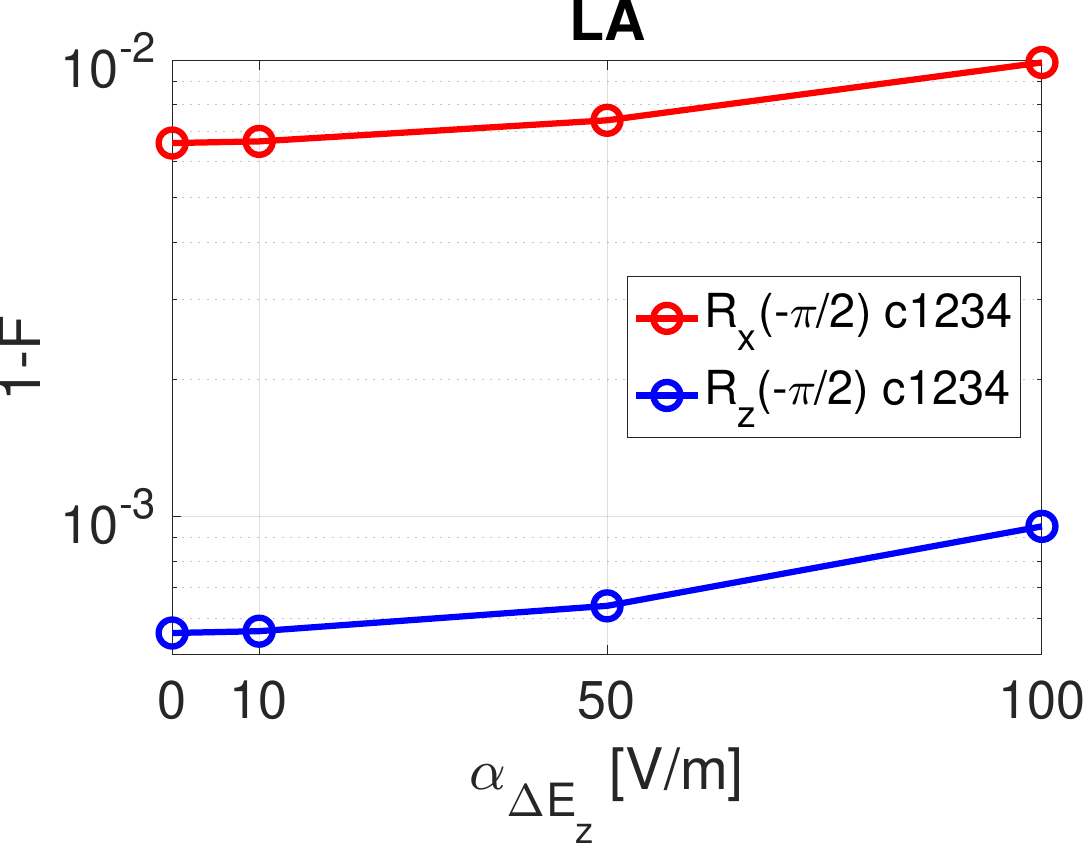}
   b)\includegraphics[width=0.3\linewidth]{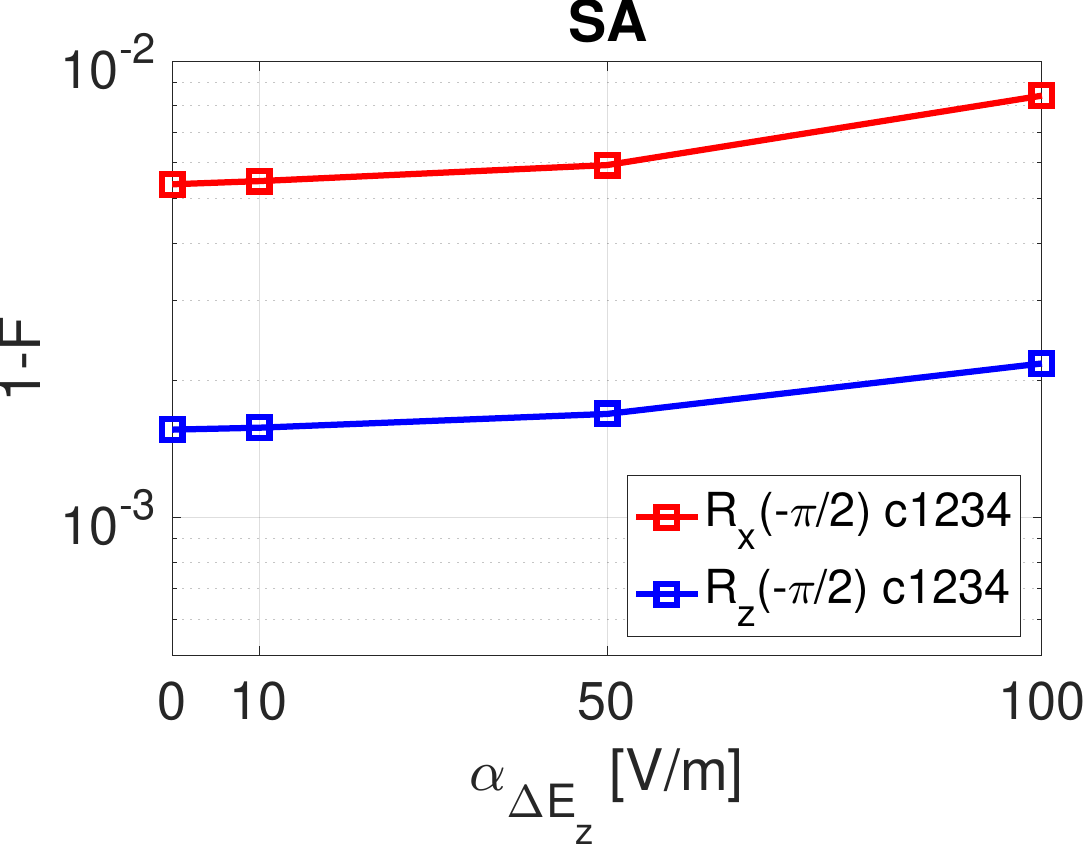} 
   c)\includegraphics[width=0.3\linewidth]{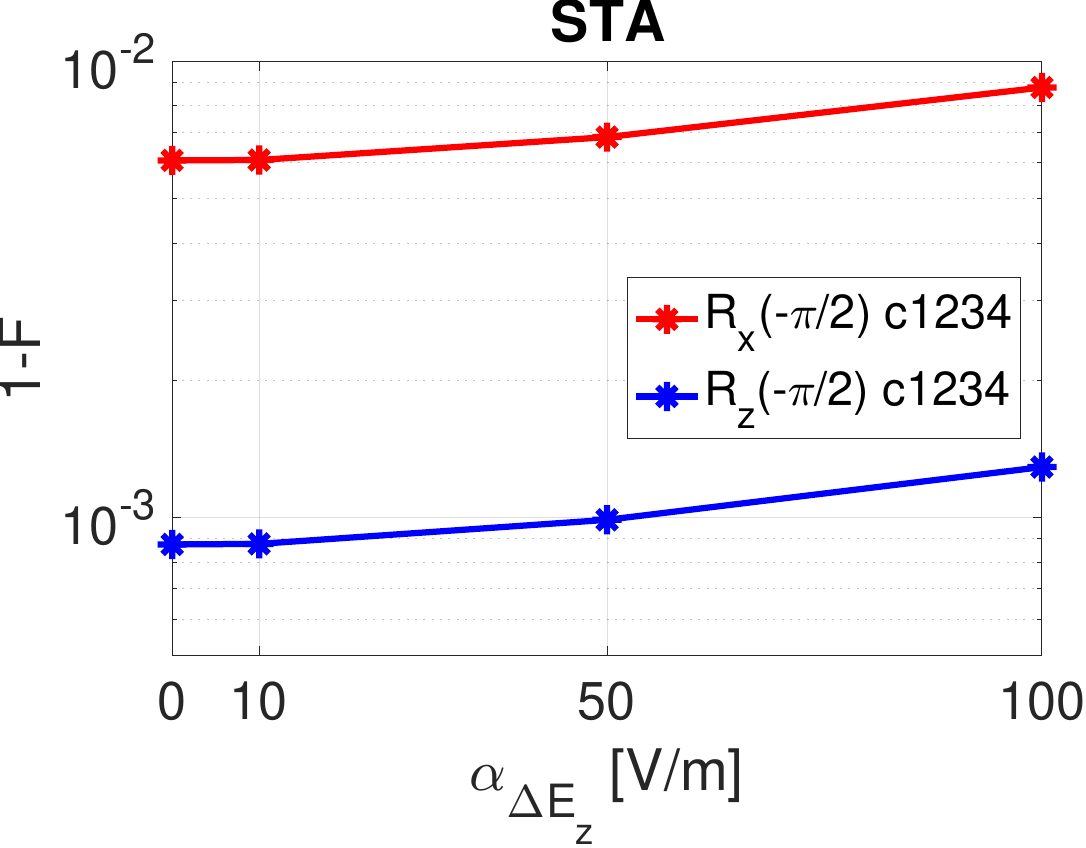}   
    \caption{Four parallel $R_{z}(-\frac{\pi}{2})$ (blue curves) and $R_{x }(-\frac{\pi}{2})$ (red curves) infidelities versus the noise amplitude $\alpha_{\Delta E_{z}}$ for the
    c1234 configuration in a) the LA, b) the SA and c) the STA.}
    \label{fig:comparisonFour1qOp}
\end{figure}

Figure \ref{fig:comparisonFour1qOp} illustrates the infidelity $1-F$ as a function of the noise amplitude $\alpha_{\Delta E_z}$ for the three arrays when four parallel one-qubit gates, $R_z(-\frac{\pi}{2})$ and $R_x(-\frac{\pi}{2})$, are applied to all qubits.

The infidelities for the four parallel one-qubit operations when $\alpha_{\Delta E_z}=$ 50 V/m are reported in Table \ref{tab:four} where the values of the only possible configuration c1234 are reported.

\begin{table}[htbp!]
\centering
\caption{Four parallel one-qubit operations infidelity values when $\alpha_{\Delta E_z}$= 50 V/m, in parenthesis the corresponding configuration is specified.} 
\resizebox{\textwidth}{!}{
    \begin{tabular}{|c|c|c|c|}
      \hline
       Four parallel  &   \multicolumn{3}{c|}{1-F} \\
       \cline{2-4}
       one-qubit operations &   LA  & SA & STA  \\       
      \hline 
       $R_z(-\frac{\pi}{2})$ &  $6.4\cdot 10^{-4}$ (c1234)& $1.7\cdot 10^{-3}$ (c1234)& $9.9\cdot 10^{-4}$ (c1234)\\ 
      \hline 
      $R_x(-\frac{\pi}{2})$   &  $7.4\cdot 10^{-3}$ (c1234) & $5.9\cdot 10^{-3}$ (c1234)& $6.8\cdot 10^{-3}$ (c1234)\\ 
      \hline
    \end{tabular}}
   
    \label{tab:four}
\end{table}

We observe that for $R_z(-\frac{\pi}{2})$, the infidelity values are very close to the case in which three parallel one-qubit operations are performed, while for $R_x(-\frac{\pi}{2})$, the infidelities are larger. The LA is the preferred geometry for $R_z(-\frac{\pi}{2})$, while the SA gives a slightly better result for $R_x(-\frac{\pi}{2})$.

\subsection{Two-qubit Operations}
\begin{figure}[htbp!]
\centering
    \includegraphics[width=0.5\linewidth]{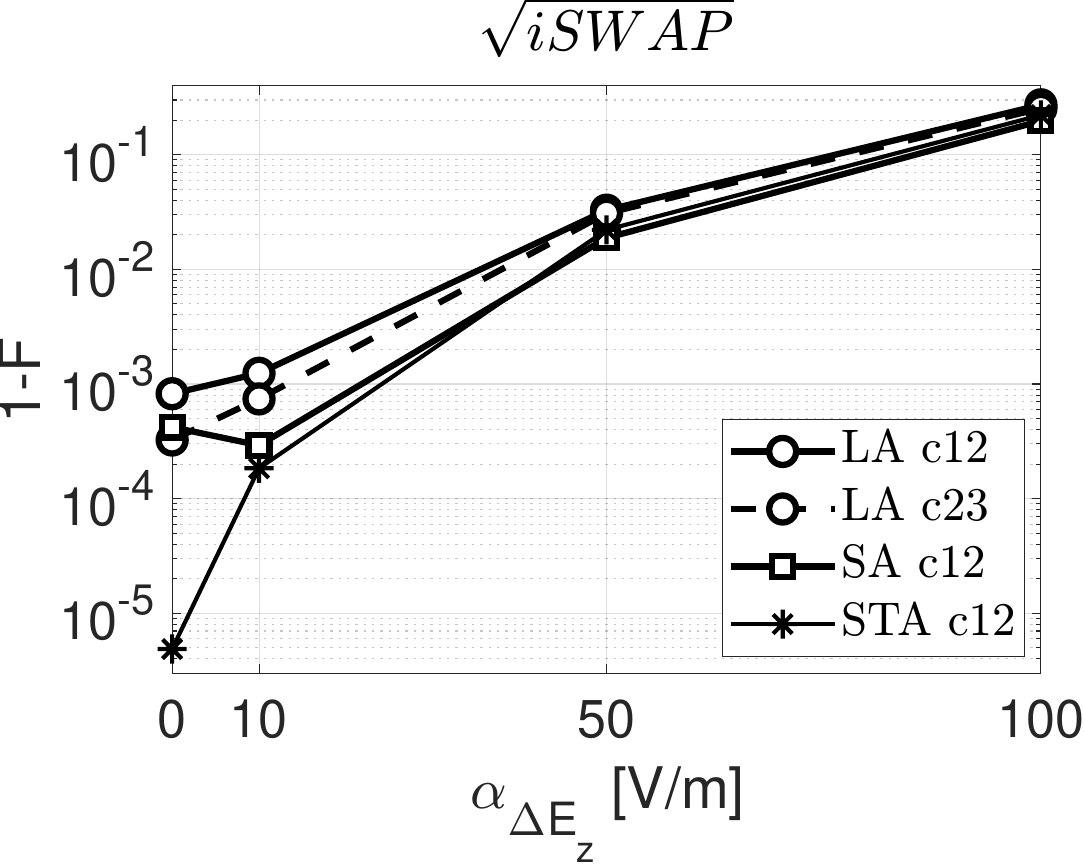} 
    \caption{$\sqrt{iSWAP}$ infidelity as a function of the noise amplitude $\alpha_{\Delta E_{z}}$ for the c12 and c23 configurations in the LA (circles), the c12 configuration in the SA (squares) and the c12 configuration in the STA (stars).}
    \label{fig:comparison2qOp_v2}
\end{figure}

Moving from one-qubit gates to two-qubit gates, Figure \ref{fig:comparison2qOp_v2} shows the $1-F$ of the two-qubit operation $\sqrt{iSWAP}$ as a function of the noise amplitude $\alpha_{\Delta E_z}$, with the other two qubits in an idle state. This analysis is conducted for the c12 and c23 configurations in LA, the c12 configuration in SA and in STA. Only the configurations with an inter-qubit distance equal to $r_0$ are considered for the $\sqrt{iSWAP}$ gate.
As expected, the infidelities of the $\sqrt{iSWAP}$ gate increase as the noise amplitude is incremented, with the different curves assuming closer values as $\alpha_{\Delta E_z}$ increases. 

The infidelities for the $\sqrt{iSWAP}$ at $\alpha_{\Delta E_z}=$ 50 V/m are reported in Table \ref{tab:2op}.

\begin{table}[htbp!]
\centering
    \caption{$\sqrt{iSWAP}$ infidelity values when $\alpha_{\Delta E_z}$= 50 V/m, in parenthesis the corresponding configuration is specified.}
    \begin{tabular}{|c|c|c|c|}
      \hline
       Two-qubit  &   \multicolumn{3}{c|}{1-F} \\
       \cline{2-4}      
           operation      &   LA  & SA & STA  \\
      \hline 
       $\sqrt{iSWAP}$ &  $3.1\cdot 10^{-2}$ (c23)& $1.9\cdot 10^{-2}$ (c12)& $2.2\cdot 10^{-2}$ (c12)\\ 
      \hline 
    \end{tabular}
    \label{tab:2op}
\end{table}

The LA is the only array that allows two different configurations, with the lowest value corresponding to c23. The SA provides the best result at this $\alpha_{\Delta E_z}$ value.

\subsubsection{Two Parallel Two-qubit Operations}
\begin{figure}[htbp!]
\centering
    \includegraphics[width=0.5\linewidth]{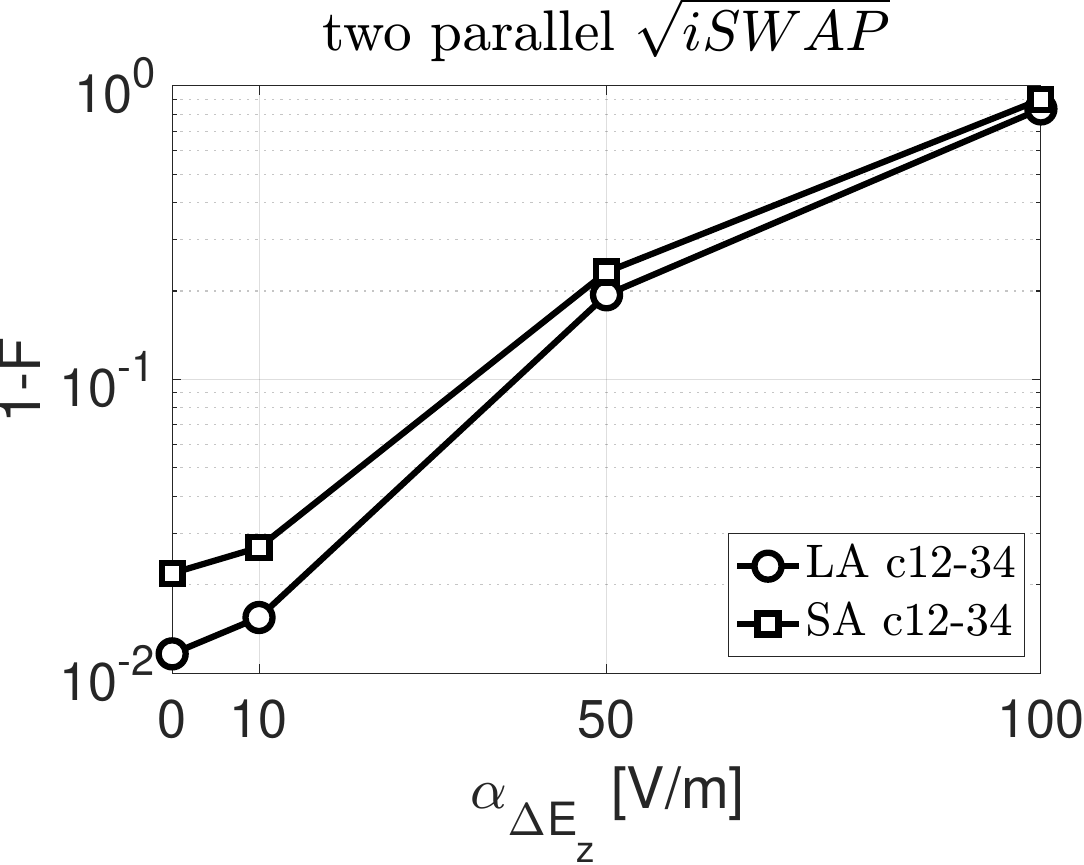}
    \caption{Two parallel $\sqrt{iSWAP}$ infidelities as a function of the noise amplitude $\alpha_{\Delta E_{z}}$ for the LA (circles) and the SA (squares) in the c12-34 configuration.}
    \label{fig:comparisonTwo2qOp_v2}
\end{figure}

Figure \ref{fig:comparisonTwo2qOp_v2} displays the $1-F$ of two parallel $\sqrt{iSWAP}$ gates compared to the noise amplitude $\alpha_{\Delta E_z}$ for the c12-34 configuration in the LA and in the SA. No results are provided for the STA because in this array it is not feasible to apply two parallel two-qubit operations with an inter-qubit distance equal to $r_0$.
As the noise amplitude increases, the infidelities of two parallel $\sqrt{iSWAP}$ gates also increase. Specifically, the LA exhibits lower infidelity compared to the SA configuration.

The resulting infidelity values when $\alpha_{\Delta E_z}$ is set to 50 V/m  are reported in Table \ref{tab:2p2op}.

\begin{table}[htbp!]
\centering
    \caption{$\sqrt{iSWAP}$ infidelity values when $\alpha_{\Delta E_z}$= 50 V/m, in parenthesis the corresponding configuration is specified.}
    \begin{tabular}{|c|c|c|c|}
      \hline
      Two parallel &   \multicolumn{2}{c|}{1-F} \\
       \cline{2-3}
      two-qubit operations &   LA  & SA   \\
      \hline 
       $\sqrt{iSWAP}$ &  $1.9\cdot 10^{-1}$ (c12-34)& $2.3\cdot 10^{-1}$ (c12-34)\\ 
      \hline 
    \end{tabular}
    \label{tab:2p2op}
\end{table}

We observe high infidelity values in both cases, but the LA performs better than the SA.

\section{Discussion}
\begin{figure}[htbp!]
\centering
    a)\includegraphics[width=0.3\linewidth]{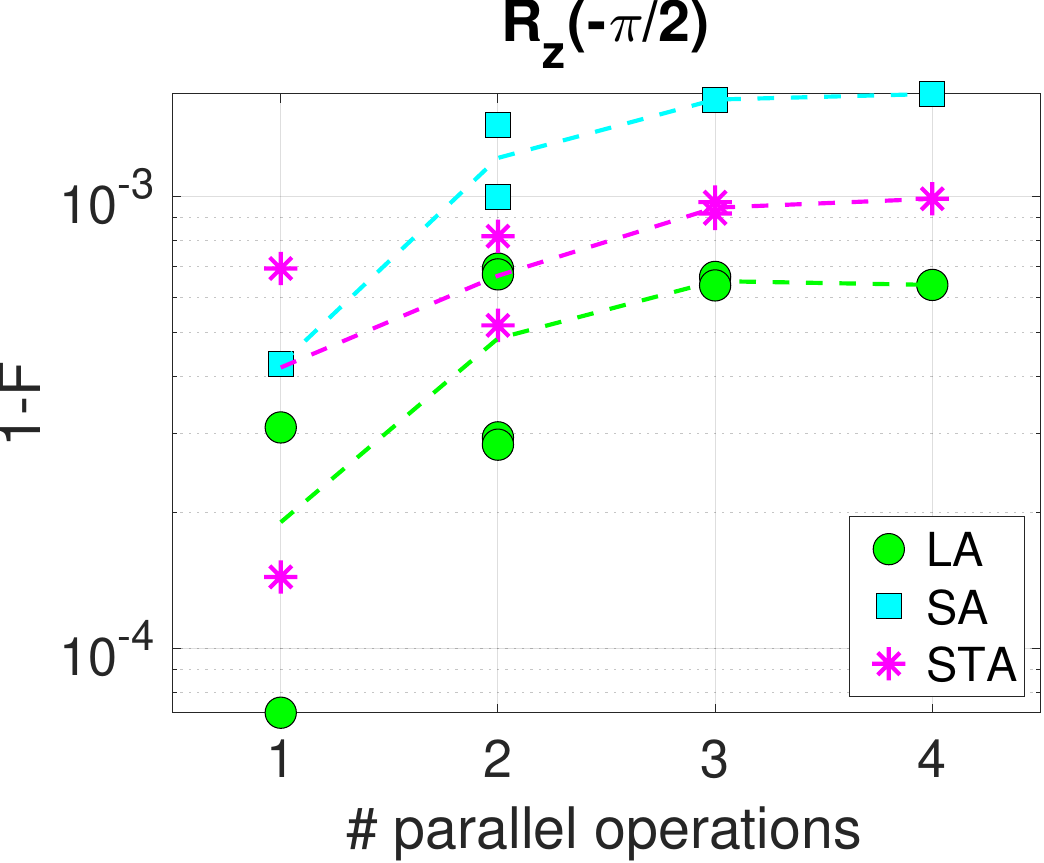}
    b)\includegraphics[width=0.3\linewidth]{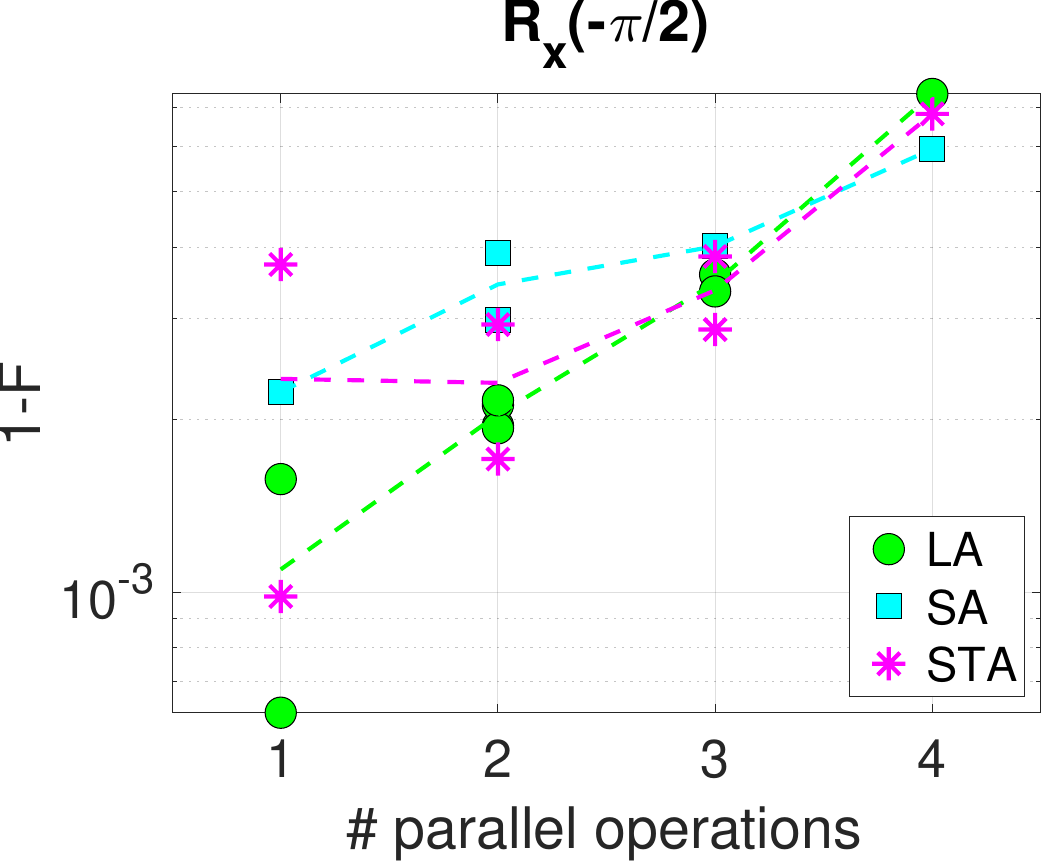} 
    c)\includegraphics[width=0.3\linewidth]{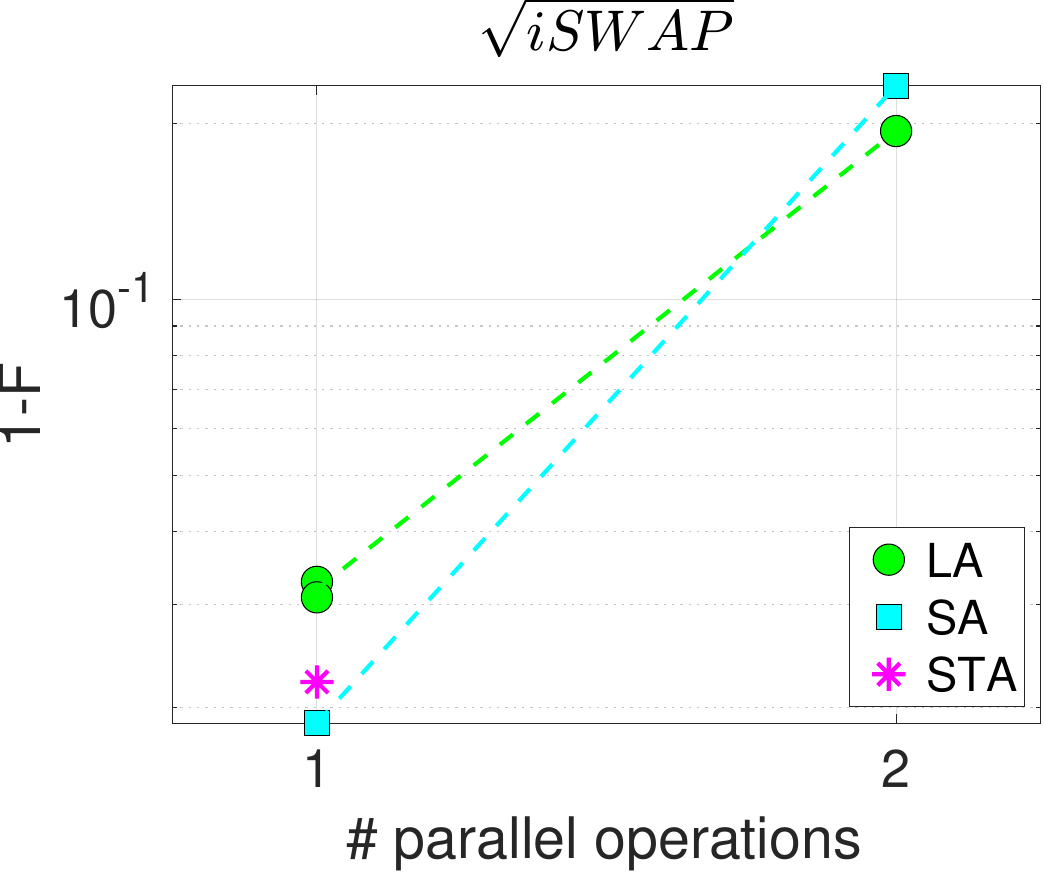}   
    \caption{Infidelity of all configurations in the LA (green circles), SA (cyan squares) and STA (magenta stars) as a function of the number of parallel a) $R_z(-\frac{\pi}{2})$ b) $R_x(-\frac{\pi}{2})$  and c) $\sqrt{iSWAP}$ gate(s) when $\alpha_{\Delta E_z}$=50 V/m. The dashed lines connect points representing the average infidelity values over different configurations for each number of parallel operations.}
    \label{fig:comparisonInF-numParOp_gates_alpha50}
\end{figure}

In Figure \ref{fig:comparisonInF-numParOp_gates_alpha50}, the simulated infidelity of each considered gate applied to all the configuration in the three different arrays is reported as a function of the number of applied parallel operations.

Figure \ref{fig:comparisonInF-numParOp_gates_alpha50}(a) displays the infidelity results of $R_z(-\frac{\pi}{2})$, demonstrating significant variability between different configurations in different arrays for one and two parallel operations. Moreover, the infidelity saturates as the number of parallel operations exceeds two, regardless of the array type. 

For each array, the plateau is effectively highlighted by the dashed line passing through points obtained by averaging the infidelities over different configurations with the same number of parallel operations. Figure \ref{fig:comparisonInF-numParOp_gates_alpha50}(a) shows that the average infidelity in LA (1-$F$= 6.4$\cdot 10^{-4}$) is lower than that in STA (1-$F$= 9.9$\cdot 10^{-4}$), which in turn is lower than that in SA (1-$F$= 1.7$\cdot 10^{-3}$).
This ordering of infidelity plateaus corresponds to the one provided by analyzing the parameter $\delta$ in the three different arrays, that gives an estimation of the interaction density and is defined here as: 
\begin{equation} \label{eq:delta}
\delta \equiv \sum_{i=1}^{N-1} \sum_{j=2,j > i}^{N} f(r_{ij}),
\end{equation}
where $f(r_{ij})$ is a function of the inter-qubit distance $r_{ij}$ between qubit $i$ and qubit $j$. Given that each array of FF qubits has an interacting Hamiltonian $\hat{H}_{int}$ proportional to $r_{ij}^{-3}$ (see Equation 6 in \cite{DeMichielis-AQT-2024}) it is natural to choose $f(r_{ij}) = r_{ij}^{-3}$. The resulting $\delta$ values for the three arrays considered are $\delta_{LA}$=3.29 $r_{0}^{-3}$, 
$\delta_{STA}$=3.51 $r_{0}^{-3}$, and 
$\delta_{SA}$=4.70 $r_{0}^{-3}$.
Thus $\delta_{LA}<\delta_{STA}<\delta_{SA}$.

In Figure \ref{fig:comparisonInF-numParOp_gates_alpha50}(b), the infidelity of $R_x(-\frac{\pi}{2})$ similarly exhibits significant variability among different configurations in different arrays for a single operation, while the variability decreases as the number of parallel operation increases. The average infidelity strongly increases with the number of parallel gates.
These increasing trends can be qualitatively explained by considering that the addition of parallel control signals, which modulate the energy of each qubit, also modifies the Electric Dipole Spin Resonance (EDSR) frequency of each qubit due to unwanted qubit mutual coupling. Since the AC control signal frequency applied to each qubit in the array is set to a constant value corresponding to the EDSR frequency of an isolated FF qubit, the qubits are progressively driven further out of resonance as the number of parallel operations applied to the array increases. This causes an infidelity increase of the entire operation which, however, does not exceed $8 \cdot 10^{-3}$ for four parallel operations. 

It is worth pointing out that in Figures \ref{fig:comparisonInF-numParOp_gates_alpha50}(a) and (b), the STA configurations with the highest infidelities - namely c2, c12, c123 and c1234 for increasing number of parallel operations - include the central qubit, denoted by index 2. 
The central qubit has the highest interaction with the other three noisy qubits, being distant $r_0$ from each one of the three qubits, a condition not met by the other qubits in STA. This unwanted coupling effect on the central qubit is particularly evident when focusing on a single operation, where the STA configuration c2 yields the worst results among the different arrays.

In Figure \ref{fig:comparisonInF-numParOp_gates_alpha50}(c) the infidelity of $\sqrt{iSWAP}$ shows less significant variability between different configurations in different arrays for both one and two parallel two-qubit operations than one-qubit gate infidelities. 
The average $\sqrt{iSWAP}$ infidelities increase as the number of parallel gates is increased, ranging from 0.02 for one operation to 0.2 for two parallel ones. 
This confirms that two-qubit gates are likely to be the limiting factor for the successful execution of an algorithm and poses a roadblock in the exploitation of parallel two-qubit gating in such small arrays of FF qubits.

\section{Conclusions} 
The effects of parallel gating on the entanglement gate fidelity have been simulated in linear, square, and star arrays of four FF qubits affected by 1/f noise.

Upon comparing the fidelity results among the three arrays at a reasonable noise level, we observed a lower average $R_z(-\pi/2)$ infidelity in the linear array compared to the star and square arrays. The infidelity values saturate as the number of parallel operations exceeds two. 
For $R_x(-\pi/2)$, the infidelities are higher than those for $R_z(-\pi/2)$, but the differences in average infidelities among the arrays decreases as the parallelism of $R_x(-\pi/2)$ increases. A similar trend is observed for $\sqrt{iSWAP}$ gates, which exhibit the highest infidelities. Parallel $\sqrt{iSWAP}$ gates result in very high infidelities suggesting that parallelization of two-qubit operations should be avoided in such small arrays.

The infidelity results of each gate in each configuration versus the number of parallel operations reported in this study provide critical information for a  quantum compiler. This information helps in selecting the optimal quantum circuit, including the correct sequence of gates with the best depth and parallelism, to implement the desired quantum algorithm with the lowest infidelity.

\medskip
\section*{Acknowledgements} 
The work was partially funded by PNRR MUR projects PE0000023-NQSTI and CN0000013-HPC financed by the European Union – Next Generation EU.

\bibliography{Ref}
\bibliographystyle{unsrt}

\end{document}